\documentclass{article}
\usepackage{arxiv}
\usepackage[numbers]{natbib}
\usepackage[utf8]{inputenc} 
\usepackage[T1]{fontenc}    
\usepackage{hyperref}       
\usepackage{url}            
\usepackage{booktabs}       
\usepackage{amsfonts}       
\usepackage{nicefrac}       
\usepackage{microtype}      
\usepackage{color,graphicx}
\usepackage{lipsum}
\usepackage{soul}
\usepackage{amsmath}

\title{Comments on the symmetry breaking condition in MacDowell-Mansouri action.}

\author{I. D\'iaz-Salda\~na, \\ 
Departamento  de F\'{\i}sica de la Universidad de Guanajuato \\
A.P. E-143, C.P. 37150, Le\'on, Guanajuato, M\'exico\\
\texttt{ ai.diazsaldana@ugto.mx} 
\And
J. C. L\'opez-Dom\'inguez\\
Unidad Acad\'emica de F\'isica, Universidad Aut\'onoma de Zacatecas,\\
Calzada Solidaridad esquina con Paseo La Bufa S/N, C.P. 98060, Zacatecas, Zacatecas., M\'exico.\\
 \texttt{jlopez@fisica.uaz.edu.mx}
\And
M. Sabido \\ 
Departamento  de F\'{\i}sica de la Universidad de Guanajuato \\
A.P. E-143, C.P. 37150, Le\'on, Guanajuato, M\'exico\\
\texttt{msabido@fisica.ugto.mx} 
\And
  J. E. Rosales-Quintero \\
  Preparatoria 2 de octubre de 1968, Universidad Aut\'onoma de Puebla,\\ P.O. Box
1364, 72000 Puebla, México\\
  \texttt{ jeduardo.rosales@correo.buap.mx, jose.rosalesquintero@viep.com.mx}
}

\begin{document}
\maketitle

\begin{abstract} 
In this work we study the symmetry breaking conditions, given by a (anti)de Sitter-valued vector field, of a full  (anti)de Sitter-invariant  MacDowell-Mansouri inspired action.  We show that under these conditions the action breaks down to   General Relativity with a cosmological constant, the four dimensional  topological invariants, as well as the Holst term. We obtain the equations of motion of this action, and analyze the symmetry breaking conditions. 
\end{abstract}

\keywords{Gauge Theories of Gravity; Torsion.}
\section{INTRODUCTION}
The greatest example of the geometrization of the fundamental interactions is Einstein's General Relativity (GR). 
{Since the formulation of GR, gravity has gone hand in hand with geometry. Currently} the
 gravitational interaction is identified as a manifestation of a curved space-time. The description of the remaining fundamental interactions is also described by the geometrical theory known as Yang-Mills (YM) theory. Although there  are many differences between GR and YM, there have been attempts to unify them in the framework of classical field theory \cite{Peldan1990,Krasnov2018}.
Different approaches have been proposed to construct a unified model of the fundamental interactions. For example, one can consider higher dimensional models of gravity, where the metric is a fundamental field from which
the particle physics interactions are constructed \cite{farlie_1,farlie}. On the other hand, in YM the interactions are described  by a connection associated with a symmetry group in a non-dynamical space-time {(in contrast to GR).  Fortunately,  there are several proposals for YM type gauge theories of gravity. These formulations that are known in the literature as pure connection actions for gravity \cite{Capovilla1991,Krasnov2011,Rosales-Quintero2016,Mitsou2019,Rosales-Quintero2020}. In these descriptions of gravity, the fundamental field is the  gauge field of a symmetry group $G$. The metric is not longer fundamental, it is a derived object, hence GR is obtained from the proposed gauge theory.}\\
In the 70's MacDowell and Mansouri (MM) \cite{MacDowell1977,Lopez-Dominguez2018,Obregon2012} were successful in finding  a gauge theory for gravity. In the MM formulation,  the $SO(1,4)$ symmetry is spontaneously broken  to $SO(1,3)$  to reproduce the  four-dimensional geometry of gravity. This is achieved by considering that the Lie algebra is $\mathfrak{so(1,4)}=\mathfrak{so(1,3)}\oplus \mathbb{R}^{1,3} $ is a reductive algebra \cite{Wise2009,Wise2010}. Then the Cartan-Killing form can be ``cut off'' in such a way that we keep the $\mathfrak{so(1,3)}$  part of the full $\mathfrak{so(1,4)}$  valued fields. In the same spirit, Stelle and West \cite{Stelle1979,Stelle1980} considered an  action where a symmetry  breaking mechanism  is induced by introducing a field $v$. This field represents the coordinates of a point in an internal anti-de Sitter space, where the motions induced by parallel transport takes place.  The gravitational vierbein and spin connection are identified from the original SO(1,4) gauge fields. Consequently, by the  symmetry mechanism, Einstein-Cartan theory plus the Euler class are obtained \cite{Lopez-Dominguez2018}. A self-dual spin connection approach for MM has been given by \cite{Nieto1996,Ramirez1998}, where duality is defined with respect to the corresponding internal symmetry group indices. Also, a formal derivation of the MM formulation from a topological $BF$ theory plus a regular interaction term can be found in the literature (see \cite{Freidel:2005ak} and references therein). In this work the dynamical sector  leads to Einstein-Cartan gravity with the cosmological term plus the Holst term. The topological sector is given by the Nieh-Yan, Pontryagin, and Euler classes.\\
The paper is arranged as follows. In section II, we present a brief account of the Stelle-West  symmetry breaking  approach for the Palatini action with cosmological constant term  and Euler class. In section III, we generalize a Stelle-West formulation, which besides  the Palatini action has the cosmological constant term. We also get the Holst modification with the Immirzi parameter and three topological terms, Nieh-Yan form, Euler class and  $SO(1,4)$ Pontryagin term. In section IV, without explicitly breaking the symmetry, we give two conditions for obtaining gravity. The first condition yields Einstein's equations plus the cosmological constant term and torsionless equation. The second condition, allows us to construct a family of topological torsionless field theories. Section V is devoted to discussion and outlook.\\
{We will like to finish this section by clarifying some notation}. We have labeled $\mathfrak{so(1,4)}$ Lie algebra indices by capital Latin letters $\{ A, B, C, \ldots \}$,   $\mathfrak{so(1,3)}$ Lie algebra indices by the
beginning of the Latin alphabet lowercase letters $\{a, b, c, \ldots \}$ and Greek alphabet letters for space-time indices $\{ \mu, \nu, \rho, \ldots \}$. Also, we consider the Minkowski metric as $\eta_{ab}=diag(-1, 1, 1, 1)$ and $\eta_{AB}=diag(-1, 1, 1, 1, 1)$, and we establish $\eta_{AB, CD}=\eta_{AC}\eta_{BD}-\eta_{AD}\eta_{BC}$ and $\eta_{ab,cd}=\eta_{ac}\eta_{bd}-\eta_{ad}\eta_{bc}$. In addition, we have defined the totally anti-symmetric tensor $\epsilon$ with $\epsilon^{01234} =+1$ and $\epsilon_{01234} =-1$ and,  in four dimensions,  $\epsilon^{0123} =+1$ and $\epsilon_{0123} =-1$. Finally, we define $G_{(ab)} = G_{ab} + G_{ba}$ and $G_{[ab]} = G_{ab} - G_{ba}$.


\section{MacDowell-Mansouri action by Stelle-West Approach}
Let us briefly describe the MacDowell-Mansouri (MM) approach {to gravity}. MM theory is a gauge theory of gravity with the gauge group $G \supset SO(3, 1)$, where $G$ depends on the sign of the cosmological constant $\Lambda$. In this work, we will
consider the case when $\Lambda>0$ and use $SO(1,4)$ (the case with $\Lambda<0$  
can be calculated straightforward). {Moreover, space-time is a 4-dimensional oriented smooth manifold} 
$\mathcal{M}=\mathbb{R}\times \Sigma$.  $\Sigma$ is  a compact manifold  without boundary and $\mathbb{R}$ represents the evolution parameter. Now we choose a principal $SO(1,4)$-bundle $P$ over $M$. Because MM theory is essentially a pure connection action of gravity \cite{Krasnov2018,Wise2010}, the fundamental field for constructing the action is the connection $A$. The connection $A$ is an $\mathfrak{so(1,4)}$ valued 1-form on $\mathcal{M}$. It can be written as  $A=A^{\ AB}_{\mu}t_{AB}dx^{\mu}$, where $t_{AB}$ are the skew-symmetric  generators of the Lie algebra $\mathfrak{so(1,4)}$ and satisfy
\begin{equation}
[t_{AB},t_{CD}]=f_{ABCD}^{\ \ \ \ \ \ \ \ \ EF} \ t_{EF}=\frac{1}{4}\
\eta_{AB,[C}^{\ \ \ \ \ \ \ \ [E }\eta_{D]}^{\ \ F]} t_{EF}.
\end{equation}
{Because} $\mathfrak{so(1,4)}$ is a reductive geometry \cite{Wise2010,Wise2009}, {we have} $\mathfrak{so(1,4)}\cong \mathfrak{so(1,3)} \oplus \mathbb{R}^{1,3}$. Then, we can write the gauge field as follows
\begin{equation}     \label{eq: Decomposition Connection}
A_{\mu}^{\ AB}= \left(\begin{array}{cc}
    A_{\mu}^{\ ab}  & A_{\mu}^{\ a4} \\
    A_{\mu}^{\ 4b}  & 0
\end{array}\right)=
\left(\begin{array}{cc}
  \omega_{\mu}^{\ ab}     & -\frac{1}{l}\ e_{\mu}^{\ a} \\
   \frac{1}{l}\ e_{\mu}^{\ b} & 0
\end{array}\right),
\end{equation}
{where we  identify $\omega_{\mu}^{\ ab}$  with the spin connection and  $e_{\mu}^{\ a}$ with the tetrad field. Also, $l$ a is constant with length dimension introduced for units requirement}. The
covariant derivative $\boldsymbol{\mathcal{D}}$ of the de Sitter group acts over Lie algebra valued fields $\xi=\xi^{AB}t_{AB}$ as follows
\begin{equation}   \label{Covariant derivative}
\boldsymbol{\mathcal{D}}\xi=\left[ D\xi^{ab}+\frac{1}{l}\left( e^{a}\wedge
\xi^{b4}-e^{b}\wedge \xi^{a4} \right) \right]t_{ab} +2\left[
D\xi^{a4}-\frac{1}{l}\ e^{b}\wedge \xi_{ b}^{\ a} \right]t_{a4},
\end{equation}
and we have introduced the Lorentz covariant derivative $
D\chi^{ab}=d\chi^{ab}+\omega^{ac}\wedge \chi_{c}^{\
b}+\omega^{bc}\wedge \chi_{ \ c}^{ a}$. \\
The field strength is defined as $F=dA+\frac{1}{2}\
[A,A]$, with
\begin{equation}
F^{ AB}= \left(\begin{array}{cc}
    F^{ ab} & F^{a4} \\
    F^{ 4b}  & 0
\end{array}\right)=
\left(\begin{array}{cc}
    R^{ab}-\frac{1}{l^2}\ \Sigma^{ab}            & -\frac{1}{l}\ T^{a} \\
  \frac{1}{l}\ T^{b}                             & 0
\end{array}\right),
\end{equation}
{where $T^{a}=De^{a} = de^{a}+\omega^{a}_{\ c}\wedge e^{c}$ is the torsion and $R^{ab} = d\omega^{ab}+\omega^{ac}\wedge\omega_{c}^{\ b}$  is the curvature. The torsion and the curvature are $\mathfrak{so(1,3)}$ valued two-forms. We  have also  defined $\Sigma^{ab} =e^{a}\wedge e^{b}$}.\\
Stelle and West considered the action \cite{Stelle1979,Stelle1980}
\begin{equation}     \label{Eq: UsualMM}
I_{MM}\left[ A, v \right]=\frac{1}{\beta_{1}}\  \int_{\mathcal{M}} \
\epsilon^{ABCDE}F_{AB}\wedge F_{CD} v_{E} + \rho
(v^{E}v_{E}-l^{-2}),
\end{equation}
{where $\beta_{1}=4Gl^{-3}$} \footnote{We work in units where $c$,  $16\pi\hbar=1$ and $16\pi G\hbar=l_{p}$.}, $v$ is a vector field with dimension of length, $\rho$ is four-form acting as a
Lagrange multiplier and $l$ is  the same real constant in Eq.(\ref{eq: Decomposition Connection}). The value of $\beta_{1}$ was chosen in accordance with \cite{Freidel:2005ak}. To obtain GR we break the $SO(1,4)$ symmetry down to $SO(1,3)$. This is achieved  by choosing a preferred direction for the vector field  $v^{E}$ and imposing that in the resulting subgroup $v$ remains fixed.  We will refer to these conditions as the symmetry breaking conditions (SBC). 
{Taking $v^{a}=0$ in $v^{E}=(v^{a}, v^{4})$  and imposing the field equation for $\rho$, we get $v^{4}=1/l$. Then  the action in Eq.(\ref{Eq: UsualMM}) takes the form}
\begin{eqnarray}  \label{eq: MM original GR+EE}
I_{MM}\left[ A  \right]&=& \frac{1}{\beta_{1}}\ \int_{\mathcal{M}} \
\epsilon^{abcd}F_{ab}\wedge F_{cd}\
\frac{1}{l}\\
&=&\ \frac{1}{l\beta_{1}}\ \int_{\mathcal{M}} \ \epsilon^{abcd}\left(
R_{ab}\wedge R_{cd}-\frac{2}{l^2}\ R_{ab}\wedge
\Sigma_{cd}+\frac{1}{l^4}\ \Sigma_{ab} \wedge \Sigma_{cd}\right).\nonumber
\end{eqnarray}
From this equation, we recognize the Euler class plus the Palatini action with non-vanishing cosmological constant. The cosmological constant is given by  $\big(\frac{1}{l^2}=\frac{ \Lambda}{3}\big)$. 

\section{A Stelle-West generalization}
In this section we propose a generalization to the action in Eq.(\ref{Eq: UsualMM}) that not only includes the terms in Eq.(\ref{eq: MM original GR+EE}), but also the Nieh-Yan term and the second Chern class. Furthermore, we want to construct the action in such a way that it can be identified with a  pure connection self-dual formulation of gravity. The construction is inspired by the Pleba\'nski formulation \cite{plebanski} and the Capovilla, Dell and Jacobson formulation \cite{Capovilla1991}.
We will start by defining a pseudo-projector for $\mathfrak{so(1,4)}$ as follows
\begin{eqnarray}  \label{eq: Pseudoprojector}
\Pi^{ABCD}\xi_{CD}&=&\frac{1}{2}\ \bigg(\alpha \eta^{AB, CD}+
\epsilon^{ABCDE}v_{E} \bigg)\xi_{CD}\nonumber\\
&=&\frac{1}{2}\ \bigg(\alpha \eta^{AB, CD}+
\widetilde{\epsilon}^{ABCD} \bigg)\xi_{CD}=\widetilde{\xi}^{AB},
\end{eqnarray}
where $\alpha$ is a constant with units of $length^{-1}$, that will be related to the Immirzi parameter and  the cosmological constant. 
The proposed  action is
\begin{equation}\label{Eq: MM Fundamental}
I_{GMM}\left[ A, v \right]= \frac{1}{\beta_{2}}\int_{\mathcal{M}} \\ 
\prec \widetilde{F} \wedge \widetilde{F} \succ+\rho(v^{E}v_{E}-l^{-2}),
\end{equation}
where $\prec \quad , \quad \succ$ is the trace over the Lie algebra generators and $\beta_{2}=4\gamma Gl^{-4}$. We can see that Eq.(\ref{Eq: MM Fundamental}) is a pure connection YM type action and the metric is not  explicitly in the action. The action $I_{GMM}$, can be thought as the integral of the Pontryagin density ``twisted'' by the pseudoprojector \footnote{In this work, the have defined a pseudoprojector $\Pi$ as a projector operator that maps a  Lie algebra bi-vector field onto a  Lie algebra bi-vector field, and it does not satisfy $\Pi\Pi=\Pi$.}.
{Interestingly}, if instead of taking $FF$ product we consider a $BF$ formulation in Eq.(\ref{Eq: MM Fundamental}), the action gives rise to Einstein conformally flat spaces \cite{Besse1987}. \\
 Using Eq.(\ref{eq: Pseudoprojector}), the action  can be rewritten as \footnote{ The last action is related to the model that was proposed in \cite{Westman} where the action is constructed by hand. In that work, the consider the connection one-form $A$ and  $v$ as the fundamental fields, combining the two and looking for a consistent action (invariant under the gauge group and polynomial in the basic fields). In our case  the action is  gauge invariant and is derived from an Yang-Mills type action.}
\begin{eqnarray}     \label{Eq: MM1}
I_{GMM}&=& \frac{1}{\beta_{2}} \int_{\mathcal{M}} \ (\alpha^2-v^{E}v_{E})F^{AB}\wedge F_{AB}+
\alpha\widetilde{\epsilon}^{ABCD}F_{AB}\wedge F_{CD} \nonumber\\
&-&2F^{AC}\wedge
F_{C}^{\ B}v_{A}v_{B}+ \rho (v^{E}v_{E}-l^{-2}).
\end{eqnarray}
Using the field  equation for $\rho$, together with the SBC  and the vector $v$ {with vanishing components $v^{a}$}, the action reads
\begin{displaymath}
I_{GMM}\left[ e,\omega \right] = \frac{6}{G\Lambda}\ \int_{\mathcal{M}} \
\frac{\gamma^2-1}{2\gamma}R^{ab}\wedge R_{ab}+ \frac{1}{2}\ \epsilon^{abcd}
R_{ab}\wedge R_{cd}+\frac{\Lambda\gamma}{3} \ d(e^{a} \wedge
T_{a})
\end{displaymath}
\begin{equation}
-\frac{1}{2G}\ \int_{\mathcal{M}} \ \bigg(\epsilon^{abcd}R_{ab}\wedge
\Sigma_{cd}-\frac{1}{\gamma} R^{ab}\wedge \Sigma_{ab}-\frac{\Lambda}{6}\
\epsilon^{abcd}\Sigma_{ab} \wedge \Sigma_{cd} \bigg),
\end{equation}
we have defined $\alpha=\gamma/l$, where $\gamma$ is the Barbero-Immirzi (BI) parameter.\\    
 We have separated the action in its topological and dynamical sectors. The first three terms in the action  are the second Chern class, the Euler class and the Nieh-Yan Class and all correspond to the topological sector. The last three terms,  Palatini action, the Holst term and the cosmological constant correspond to the dynamical sector. We have associated the Holst term to the dynamical sector. Although at the classical level it does not affect the dynamical behavior of the theory, we keep in mind its apparent effect at the quantum level, related to the prediction of the spectra of geometrical operators, area and volume, and consequently appearance in the black hole entropy calculation. Additionally, it labels inequivalent quantizations in the framework of kinematical loop quantum gravity and modifies the symplectic structure, which can not be unitarily implementable at quantum level. 
{In summary,  from Eq.(\ref{Eq: MM Fundamental}) and SBC} we obtain GR with cosmological constant and the Immirzi parameter plus topological terms. Therefore Eq.(\ref{Eq: MM Fundamental}) can be considered a pure connection action for gravity.\\
To construct the  metric tensor we use the covariant derivative on $v^{A}$ \cite{Stelle1979,Stelle1980} 
\begin{equation} \label{eq: Metric}
 g_{\mu\nu}= e_{\mu}^{\ a}e_{\nu a} =\boldsymbol{\mathcal{D}}_{\mu}v^{A} \boldsymbol{\mathcal{D}}_{\nu}v_{A},
\end{equation} 
where we have used the SBC and the fact that the metric tensor is defined upon a constant factor. The tetrad field is essentially given  in terms the vector field $v$, but since this vector is constant, then the metric information is encoded in the full $\mathfrak{so(1,4)}$ covariant derivative. {The covariant derivative connects the fibers on a fiber bundle through parallel transport. Then, the direction of $v$ gives   a preferred direction, in which the resulting bundle allows to construct GR.}

\section{Equations of motion without SBC}
Let us focus our attention on the following question, given Eq.(\ref{Eq: MM Fundamental}) and no preferred direction for $v$ ( i.e., no SBC): is it possible to introduce conditions over the full covariant derivative of $v$ and  obtain  GR? \\
To do so, let us first consider Eq.(\ref{Eq: MM Fundamental}) and its equations of motion. From this point forward, we will take  $v^{2}=l^{-2}$ and no longer constant. {When considering the action in Eq.(\ref{Eq: MM Fundamental}) with the field equation  for $\rho$,  $v$ might have non constant components}. Consequently, some terms that are proportional to $\mathcal{D}v$  can contribute \footnote{ This case was briefly mentioned in \cite{Westman} but was not explored.} and the action can not be written as a total derivative. 
  Since $v$ is not necessarily  a constant vector, we take its covariant derivative and find 
\begin{equation}  \label{eq: dv5 fundamental}
\boldsymbol{\mathcal{D}}(v^{A}v_{A})=\boldsymbol{\mathcal{D}}(l^{-2}) \quad \Rightarrow \quad v^{A}\boldsymbol{\mathcal{D}}v_A=0 \quad \Rightarrow \quad dv^{4}=-\frac{1}{v^{4}} v^{a}dv_{a},
\end{equation}
where we have written $v^A=(v^{a},v^{4})$ and $v^{4}\neq 0$.

The equations of motion from  $I_{GMM}$, are given by
\begin{eqnarray}
\label{eq: MM Fundamental TS} \epsilon^{ABCDE} F_{AB}\wedge F_{CD}          &=& \frac{4}{\alpha} \ F^{EA} \wedge F_{A}^{\ B}v_{B}, \\
\label{eq: MM Fundamental DS}  \epsilon^{ABCDE} F_{CD}\wedge \boldsymbol{\mathcal{D}}v_{E} &=& \frac{2}{\alpha} \ F^{C[A}\wedge
  \boldsymbol{\mathcal{D}}(v^{B]}v_{C}).
\end{eqnarray}
The first equation is related to the variation with respect to $v$, the second to the variation  respect to $A$. The  equation of motion for $A$ implies a dynamical equation for $v$. {Then}, our task is to find some
solutions for this dynamical behavior which led us to understand
the role played by $v$ in a more fundamental level. In the next subsections we will present  two different dynamical
solutions for $v$. 
\subsection{Case $\mathcal{D}v^{E}=A^{E4}v_{4}$}
Let us consider the case $\boldsymbol{\mathcal{D}} v^{E}=A^{E4}v_{4}$, with non-vanishing  $v^{4}$ for at least one component of $v$. This gauge implies the following equations
\begin{eqnarray}
\label{eq: Covariant derivative over v2.2}           v^{A}\boldsymbol{\mathcal{D}}v_{A}&=&\frac{1}{l}\ v^{a}e_{a}v^{4}=0,\\
\label{eq: Covariant derivative v different zero.2}      Dv^{a} &=& 0,\\
\label{eq: derivative v5.2}                    dv^{4}     &=&
-\frac{1}{l}\ v^{a}e_{a}.
\end{eqnarray}
From eq.(\ref{eq: Covariant derivative over v2.2}) we identify two cases. If we take $v^{4}=0$, from Eq.(\ref{eq: derivative v5.2}) we get that $v^{a}e_{a}$ vanishes as well. However, since we are considering a non-degenerate tetrad field,  $v^{a}$ also vanishes. Consequently, we conclude that $v^{E}=0$ but this is the trivial case that we are not considering. In the second case we have a  non-vanishing  $v^{4}$ (at least in some finite regions on the manifold). As in the previous case $v^{a}e_{a}=0$, hence $v^{a}=0$. Since we are  considering non-degenerate tetrad fields and considering Eq.(\ref{eq: derivative v5.2}), we conclude that $v^{4}$ is constant and equal to $l^{-1}$. Moreover, from the algebraic constraint imposed on $v\cdot v$ we conclude that this case is equivalent to considering the case with SBC.\\
Our next step is to consider equations Eq.(\ref{eq: MM Fundamental TS}) and Eq.(\ref{eq: MM Fundamental DS}), substitute  $v^{E}=(0, 0, 0, 0, l^{-1})$. Together with the condition and $\alpha=1/(l\gamma)$, it gives the following equations
\begin{eqnarray}
(\gamma \eta^{ab,cd}+\epsilon^{abcd})T_{c}\wedge e_{d} & = & 0,\\
\epsilon^{abcd}R_{bc}\wedge e_{d}-\frac{\Lambda}{3}\epsilon^{abcd}e_{b}\wedge e_{c}\wedge e_{d} & = & \frac{1}{\gamma}DT^{a},
\end{eqnarray}
where the first equation is the well-known zero torsion condition, which allows us to write the spin connection as a function of the tetrad field. The second line describes vacuum Einstein equations. This pair of equations come directly from the condition in Eq.(\ref{eq: MM Fundamental DS}). \\
Also from Eq.(\ref{eq: MM Fundamental TS}) and  Eq.(\ref{eq: MM Fundamental DS}), we obtain
\begin{equation}
    \frac{\Lambda^2}{9}=\frac{\epsilon_{abcd}R^{ab}\wedge R^{cd}}{\epsilon_{abcd}\ e^{a}\wedge e^{b}\wedge e^{c}\wedge e^{d}},
\end{equation}
which provides an algebraic equation for the cosmological constant and it can allow us to fix its value \cite{Alexander2019,Alexander2019b}. So the gauge given  in this subsection for the covariant derivative of $v$, not only gives us GR with cosmological constant, it fixes the value of the cosmological constant. In some sense, using this gauge is almost equivalent as breaking the symmetry group. It also gives more information by giving an expression for the the cosmological constant. Accordingly, we can argue that $\Lambda$ is calculated within the theory. Finally, it is possible to calculate the metric by using Eq.(\ref{eq: Metric}).


\subsection{Case $\boldsymbol{\mathcal{D}}v^{E}=0$}
This case is of particular interest since  Eq.(\ref{eq: MM Fundamental DS}) vanishes identically. Then the metric can not be constructed by considering Eq.(\ref{eq: Metric}) as in the previous cases. The gauge condition $\mathcal{D}v^{E}=0$, gives 
\begin{eqnarray}
\label{eq: Covariant derivative v different zero}      Dv^{a} &=& \frac{1}{l}\ e^{a}v^{4},\\
\label{eq: derivative v5}                    dv^{4}     &=&
-\frac{1}{l}\ v^{a}e_{a},
\end{eqnarray}
from Eq.(\ref{eq: dv5 fundamental}) and  Eq.(\ref{eq: derivative v5}), we conclude that the  tetrad field is completely defined by  the vector $v$ and its derivatives
\begin{equation} \label{eq: tetrad field as function of v}
e^{a}=\frac{l}{v^{4}}\ dv^{a}.
\end{equation}
From the last equation and Eq.(\ref{eq: Covariant derivative v different zero}) we get
\begin{equation} \label{eq: Pseudo-orthogonal condition}
\omega^{a}_{ \ b}v^{b}=0,
\end{equation}
then spin connection must be orthogonal to the Minkowski components of  $v$. By applying the $\mathfrak{so(1,4)}$ covariant derivative on the gauge condition  $\mathcal{D}v^{E}=0$, we find $F^{AB}v_{B}=0$.This expression can be decomposed in two expressions 
 \begin{eqnarray}
\label{eq: Fv identity} F^{ab}v_{b}=\bigg( R^{ab}-\frac{1}{l^2}\ \Sigma^{ab} \bigg) v_{b}& = & \frac{1}{l}\ T^{a}v_{4}\\
\label{eq: Torsion linear combination} T^{a}v_{a} & = & 0.
\end{eqnarray}
The first equation is an identity (see Appendix B) and gives no further information. 
{In the next subsections we will focus our attention on the second equation}. But before we proceed, let us point out that $v^a$ can not be constant (this can be seen from  Eq.(\ref{eq: tetrad field as function of v})).\\


\subsubsection{Zero Torsion condition.}
A solution for Eq.(\ref{eq: tetrad field as function of v}) is $T^{a}=0$. Then we can write the spin connection as function of the tetrad field and because the tetrad field depends on $v$, also $\omega$ depends on $v$
\begin{equation}
\omega=\omega(e)=\omega (v^a,v^5), \qquad \qquad e=e (v^a,v^5). 
\end{equation}
We can see from the previous equation that all the fundamental objects are given by the vector field $v^{A}$. Consequently, this vector allows  us to construct the frame bundle and (locally) the SO(1,3)-bundle, where the spin connection must be orthogonal to the vector $v^{a}$. Using Eq.(\ref{eq: Torsion linear combination}) and Eq.(\ref{eq: Pseudo-orthogonal condition}), we get
\begin{equation}   \label{eq: Fundamental for identity on EoM}
\chi^{ab}v_{b}=0, \qquad \textrm{where}\qquad \chi^{ab}= d\omega^{ab} -\frac{1}{l^{2}}\ \Sigma^{ab}.
\end{equation}
Let us note that $det|\chi|\neq 0$ is not possible, since it will imply that  $v^{a}=0$. We  can take $det|\chi|= 0$ and calculate the null vectors $v^{a}$, but this case goes beyond the scope of this work. The other possibility is to consider $\chi^{ab}=0$, this implies that  $l^{2}d\omega^{ab}=\Sigma^{ab}$. Thus, by taking $\chi^{ab}=0$ and the zero torsion condition, we observe that the field equations derived from  $v$ reduce to 
\begin{equation}
\epsilon^{abcd} F_{ab}\wedge F_{cd} =0 \qquad \Rightarrow \qquad \epsilon^{abcd}\omega_{a}^{\ f}\wedge\omega_{fb}\wedge \omega_{c}^{\ g}\wedge \omega_{gd}=0.
\end{equation}
This equation vanishes and we don't have any dynamical equations of motion. Hence, using  this gauge in conjunction  with  $\chi=0$, gives the conditions to have a topological field theory. Therefore, we conclude that $v$ parametrizes a family of topological torsionless field theories. 

\subsubsection{Orthogonal spin connection condition.}

We now consider Eq.(\ref{eq: Torsion linear combination}) with $T^{a} \neq 0$. Then, from Eq.(\ref{eq: Pseudo-orthogonal condition}) we get
\begin{equation}
    v_a d e^a=0,
\end{equation}
but this equation is an identity (see Appendix B). In this case, the equation of motion for $v$ can be treated as the zero torsion condition, since it is the only equation that relates the spin connection with the tetrad field. Therefore, we get
\begin{equation}
    \epsilon^{abcd}F_{ab}\wedge F_{cd}=0 \rightarrow F_{ab}\wedge F_{cd}+F_{bc}\wedge F_{ad}+F_{ac}\wedge F_{db}=0.
\end{equation}
Contracting it with a vector $v^{d}$ and using Eq.(\ref{eq: Fv identity}) we find
\begin{equation}
    D\bigg[ e_{a}\wedge R_{bc}+e_{b}\wedge R_{ca}+e_{c}\wedge R_{ab} -\frac{1}{l^{2}}e_{a}\wedge e_{b}\wedge e_{c} \bigg]=0.
\end{equation}
With the help of Eq.(\ref{eq: tetrad field as function of v}) it can be rewritten as
\begin{equation}\label{33}
    D\bigg[ \frac{dv_{a}}{v^{4}}\wedge R_{bc}+\frac{dv_{b}}{v^{4}}\wedge R_{ca}+\frac{dv_{c}}{v^{4}}\wedge R_{ab} -\frac{1}{l^{2}}\frac{dv_{a}}{v^{4}}\wedge \frac{dv_{b}}{v^{4}}\wedge \frac{dv_{c}}{v^{4}} \bigg]=0.
\end{equation}
{From this equation one  obtains  $\omega$  as a function of $v$  its derivatives $\omega=\omega(v,\partial v)$.  For this particular case,  Eq.(\ref{33}) plays the role of the zero torsion condition, as it relates the tetrad and the connection. Interestingly, the tetrad field and therefore the metric, is not obtained from the Einstein field equations. Instead it is a consequence the imposed gauge, and  depends on the functional form of $v$.}


\section{Conclusions and Outlook}
In this work we have considered a generalized Stelle-West formulation of gravity by means of introducing a pseudoprojector $\Pi$ acting on $\mathfrak{so(1,4)}$ Lie algebra valued field strength. This projector has the form of a complex (anti)self-dual projector of a Pleba\'nski like formulation. Moreover, as in the original Stelle-West formulation, the vector field $v$ hidden into the projector $\Pi$, plays a central role in breaking the symmetry down to $SO(1,3)$. When considering the SBC, we not only get Einstein-Cartan theory and the Euler class, we have also obtained the Holst modification, the Nieh-Yan and the Pontryagin topological terms. 
We also considered the action without the SBC. The equations of motion  are not trivial, so we  introduced  constraints on the covariant derivative of $v$. In the first case, we imposed a condition that allowed us to recover Einstein's  equations. Furthermore, from the equation  for $v$ we obtained a condition over the cosmological constant. We find that  $\Lambda$ is calculated within the theory, 
this is consistent with the result presented in \cite{Alexander2019,Alexander2019b}. It will be intriguing to compare  the relationship between the action in Eq.(\ref{Eq: MM Fundamental}) and the quasitopological principle proposed by Alexander et. al. in \cite{Alexander2019}, for their $\theta$ term and the non-constant $\Lambda$.\\
For the second case, we were able to construct topological torsionless field theories. In this case, the Minkowski part of the vector $v$ is orthogonal to the spin connection and $\chi=0$. Since the Lie algebra is a reductive algebra and the connection is an $\mathfrak{so(1,3)}$ valued field, the  vector $v$ allows us to construct the   (local)  $SO(1,3)$ bundle  and the frame bundle. Furthermore,  the tetrad field (as well as the the metric) does not come from the Einstein field equations. Instead, is obtained from the gauge  we used and from Eq.(\ref{eq: tetrad field as function of v}), we also find that the metric depends on the functional form of $v$. 
Numerical solutions for the field $v$ (under the second condition) could be used to reconstruct the metric and characterize the resulting  manifolds and their topological structures.

In this paper we have only worked the bosonic case,  it will be interesting to consider the the supersymmetric extension for $\mathcal{N} = 1$. Also higher dimensional internal symmetry group based on MM approach for pure connection formulations of gravity for real fields, inspired in \cite{Rosales-Quintero2020}, can be considered in order to obtain a  family of torsionless conformally flat Einstein manifolds. These ideas are work in progress and will be reported elsewhere.


\section*{Acknowledgements}
The authors would like to thank the anonymous referee who provided valuable comments, which helped to improve the paper. This work is  supported by   CONACYT grants  257919, 258982.  J.C.L-D is supported by UAZ-2019-37818. M.S. is suported by CIIC/2021. J. E. Rosales-Quintero is supported by ``Est\'imulos a la investigaci\'on para doctoras y doctores 2020 del  Consejo de Ciencia y Tecnolog\'ia del Estado de Puebla", he also acknowledges support from Benem\'erita Universidad Aut\'onoma de Puebla  for granting the necessary conditions to carry out this research work. 

\appendix
\section{GR action from the Stelle-West generalization.}
In this appendix we present the derivation of th GR action (including topological terms) from  Eq.(\ref{Eq: MM Fundamental}).
We start by considering  the term $\prec\widetilde{F}\wedge\widetilde{F}\succ$
\begin{eqnarray}
\prec\widetilde{F}\wedge\widetilde{F}\succ&=&\frac{1}{4}\bigg(\alpha\eta^{AB,CD}+\epsilon^{ABCDE}v_{E})F_{CD}\wedge (\alpha\eta_{AB,FG}+\epsilon_{ABFGH}v^{H}\bigg)F^{FG}\\
&=&\frac{1}{4}\bigg(4\alpha^{2} F^{AB}\wedge F_{AB}+4\alpha\epsilon^{ABCDE}v_{B}F_{AB}\wedge F_{CD}+\epsilon^{ABCDE}\epsilon_{ABFGH}v_{E}v^{H}F_{CD}\wedge F^{FG}\bigg), \nonumber
\end{eqnarray}
and use $\epsilon^{ABCDE}\epsilon_{ABFGH}=-2\Big[\delta^{CD}_{\ \ \ \ FG}\delta^{E}_{H}+\delta^{CD}_{\ \ \ \ GH}\delta^{E}_{F}+\delta^{CD}_{\ \ \ \ HF}\delta^{E}_{G}\Big]$ to get
\begin{equation} \label{Eq: FtildeFtilde in appendix A}
\prec\widetilde{F}\wedge\widetilde{F}\succ=(\alpha^{2}-v^{E}v_{E})F^{AB} \wedge F_{AB}+\alpha \epsilon^{ABCDE}v_{E}F_{AB} \wedge F_{CD}-2F_{AC} \wedge F^{CB}v^{A}v_{B}.
\end{equation}
We are imposing the conditions $v^{E}v_{E}=l^{-2}$, $v^{E}=(0, 0, 0, l^{-1})$ and $\alpha=\gamma l^{-1}$. Then for each term in the last equation we have
\begin{eqnarray}\label{36}
F^{AB} \wedge F_{AB}&=&F^{ab} \wedge F_{ab}+F^{a4}\wedge F_{a4}+F^{4a}\wedge F_{4a}=R^{ab}\wedge R_{ab}+2(\Lambda/3)d[e_{a}\wedge T^{a}],\nonumber \\
\epsilon^{ABCDE}v_{E}F_{AB} \wedge F_{CD}&=&l^{-1}\epsilon^{abcd}[R_{ab}\wedge R_{cd}-2l^{-2}R_{ab}\wedge \Sigma_{cd}+l^{-4} \Sigma_{ab}\wedge \Sigma_{cd} ],\nonumber \\
2F_{AC}\wedge F^{CB}v^{A}v_{B}&=& -2l^{-4}T_{a}\wedge T^{a}=-2l^{-4}[d(e_{a}\wedge T^{a})+R^{ab}\wedge \Sigma_{ab}],
\end{eqnarray}
where to obtain the first and the third  equations we used the Nieh-Yan invariant form $d(e_{a}\wedge T^{a})=T_{a}\wedge T^{a}-R^{ab}\wedge \Sigma_{ab}$. Finally, by substituting Eq.(\ref{Eq: FtildeFtilde in appendix A}) in Eq.(\ref{36})) and defining $l^{-2}=\Lambda/3$ we get
\begin{eqnarray}
\prec\widetilde{F}\wedge\widetilde{F}\succ&=&\frac{2\gamma\Lambda}{3} \ \left[ \bigg( \frac{\gamma^{2}-1}{2\gamma}  \bigg) R^{ab}\wedge R_{ab}+\frac{1}{2}\ \epsilon^{abcd}R_{ab}\wedge R_{cd}+\frac{\gamma\Lambda}{3}d(e_{a}\wedge T^{a})\right .\nonumber\\  
&-& \left .\frac{\Lambda}{3}\ \left( \epsilon^{abcd}R_{ab}\wedge \Sigma_{cd}-\frac{1}{\gamma}\ R^{ab}\wedge \Sigma_{ab}-\frac{\Lambda}{6}\ \epsilon^{abcd}\Sigma_{ab}\wedge \Sigma_{cd} \right)   \right].    
\end{eqnarray}

\section{Two equations that vanish identically.}
{In this section, we show the vanishing of two equations derived from  $\mathcal{D}v^{E}=0$.}

Let us start with  Eq.(\ref{eq: Fv identity}),
\begin{eqnarray}
     \left( R^{ab}-\frac{1}{l^{2}} \Sigma^{ab} \right) v_{b} &=& \left( d\omega^{ab} + \omega^{a}_{\ c}\wedge \omega^{cb} -\frac{1}{l^{2}} e^{a} \wedge e^{b}  \right) v_{b}\nonumber \\
     &=&d(\omega^{ab}v_{b})+\omega^{ab}\wedge dv_{b}-\frac{1}{l^{2}}\ e^{a}\wedge [-ldv^{4}],
\end{eqnarray}  
where we have used $e^{a}v_{a}=-ldv^{4}$ and $\omega^{ab}v_{b}=0$.\\
By considering that $v^{4}e^{a}=dv^{a}$, we can rewrite the last equation as
\begin{eqnarray}
     \omega_{ab}\wedge \bigg[ \frac{v^{4}}{l}\bigg] \ e^{b}+\frac{1}{l}\ e^{a}\wedge dv^{4} &=& \frac{v^{4}}{l}\ \omega^{ab}\wedge e_{b}+\frac{1}{l}\ [ -d(e^{a}v^{4})+de^{a} v^{4}]\nonumber\\
     &=&\frac{v^{4}}{l}\ T^{a}-\frac{1}{l}\ d(dv^{a})=\frac{v^{4}}{l}\ T^{a},
\end{eqnarray}
   So finally we get $F^{ab}v_{b}-v^{4}/l\ T^{a}=0$.

    We now turn our attention to  Eq.(\ref{eq: Torsion linear combination}),
\begin{equation}
        v_{a}de^{a}=v_{a}d\left[\frac{1}{v^{4}}\ dv^{a}\right]=-\frac{v_{a}}{2(v^{4}) ^{2}}dv^{4}\wedge d(v_{a}v^{a}).
\end{equation}
    Let us now consider the equation $v^{F}v_{F}=v^{a}v_{a}+v^{4}v_{4}=1/l^{2}$, then the last equation reads
\begin{equation}
        \frac{v_{a}}{2(v^{4}) ^{2}}dv^{4}\wedge d((v^{4}) ^{2})=\frac{1}{v^{4}}dv^{4}\wedge dv^{4}=0.
\end{equation}
so finally,   $v_{a}de^{a}=0$.


\bibliographystyle{unsrt}
\bibliography{references}

\end{document}